\documentclass[pra,superscriptaddress,amsmath,10pt,floatfix,showpacs,preprintnumbers]{revtex4}
\usepackage[mathscr]{eucal}
\usepackage{color}
\usepackage{amsmath}
\usepackage[dvips]{graphicx}
\usepackage{epsf}
\usepackage{bm}
\newcommand{\be}{\begin{equation}}

\newcommand{\ee}{\end{equation}}
\newcommand{\bea}{\begin{eqnarray}}
\newcommand{\eea}{\end{eqnarray}}

\newcommand{\de}{\partial}

\def\slr#1{\setbox0=\hbox{$#1$}           
   \dimen0=\wd0                                 
   \setbox1=\hbox{/} \dimen1=\wd1               
   \ifdim\dimen0>\dimen1                        
      \rlap{\hbox to \dimen0{\hfil/\hfil}}      
      #1                                        
   \else                                        
      \rlap{\hbox to \dimen1{\hfil$#1$\hfil}}   
      /                                         
   \fi}

\def\be{\begin{eqnarray}}
\def\ee{\end{eqnarray}}

\begin{document}

\title{Evaluating the phase diagram of   superconductors with asymmetric spin populations}
\date{\today}
\author{Massimo~Mannarelli}
\email{massimo@lns.mit.edu} \affiliation{Center for Theoretical
Physics, Massachusetts Institute of Technology, Cambridge, MA 02139}
\author{Giuseppe~Nardulli}
\email{giuseppe.nardulli@ba.infn.it}\author{Marco~Ruggieri}
\email{marco.ruggieri@ba.infn.it}  \affiliation{Universit\`a di
Bari, I-70126 Bari, Italia} \affiliation{I.N.F.N., Sezione di Bari,
I-70126 Bari, Italia}\affiliation{TIRES, Centre of Innovative
Technologies for Signal Detection and Processing, University of
Bari, I-70126 Bari, Italia}
\begin{abstract}

The phase diagram of a non-relativistic fermionic system  with
imbalanced state populations   interacting via a short-range S-wave
attractive interaction is analyzed in the mean field approximation.
We determine the energetically favored state  for different values
of the mismatch between the two Fermi spheres  in the weak and
strong coupling regime considering both homogeneous and
non-homogeneous superconductive states. We find that the homogeneous
superconductive phase persists for values of the population
imbalance that increase with increasing coupling strength. In the
strong coupling regime and for large population differences the
energetically stable homogeneous phase is characterized by one
gapless mode. We also find that the inhomogeneous superconductive
phase characterized by the condensate $\Delta({\bf x}) \sim
\Delta~\exp{(i \bf{q \cdot x})}$ is energetically favored in a range
of values of the chemical potential mismatch that shrinks to zero in
the strong coupling regime.

\end{abstract}
\preprint{MIT-CTP-3736 } \preprint{Bari-TH -535/06} \maketitle

\section{Introduction}
Fermionic superfluidity driven by  S-wave short range interaction is
a collective phenomenon characterized by formation of correlated
pairs of half-integer spin particles. This effect is relevant in
many theoretical and experimental fields. Of particular interest are
systems of trapped cold atoms of two fermionic species where the
interaction between the fermions can be varied over a wide range
employing a Feshbach resonance \cite{Feshbach}. A general feature of
such systems is that when the interaction between the fermions is
increased the systems evolve from a weak-coupled superconductive BCS
state of Cooper pairs to a Bose-Einstein condensate (BEC) of
tightly-bound diatomic molecules \cite{BEC,Melo}.

In a standard BCS superconductor the chemical potentials of the two
fermionic species are equal. A small chemical potential difference
$\delta\mu$ cannot disrupt BCS superconductivity. As a matter of
fact the superconductive state with equal number densities is
energetically favored in comparison with a normal state with a
fermionic imbalance. On the other hand, as  pointed out in
\cite{Clogston:1962} in the weak coupling regime, BCS
superconductivity cannot persist for large values of  $\delta\mu$.
Indeed there exists an upper limit for $\delta\mu$ (the so-called
Clogston limit), beyond which the homogeneous superconductive state
is no longer energetically favored.

Systems with  unequal populations of two fermionic species have been
widely studied
\cite{Clogston:1962,LOFF,Sarma,Muther:2002mc,Bedaque:2003hi,
Liu:2002gi,Gubankova:2003uj,Forbes:2004cr,Carlson:2005kg,Castorina:2005kg,Pao,Son:2005qx,
Sheehy,Yang:2005,trap,Chen,Bulgac:2006gh,Yang:2006ez,Gubankova:2006gj}.
For values of $\delta\mu$  larger than the Clogston limit several
non trivial phases have been proposed. They  include the Phase
Separation (PS) state \cite{Bedaque:2003hi}, the Breached Pair (BP)
superfluidity \cite{Liu:2002gi,Gubankova:2003uj,Forbes:2004cr}, the
Deformed Fermi sea pairing (DFSP) \cite{Muther:2002mc} and the
non-homogeneous LOFF \cite{LOFF} phase (see \cite{Casalbuoni:2003wh}
for a review).

Recent experiments with trapped cold atoms \cite{ketterle,Partridge}
indicate that  both the BCS and the BEC states can sustain large
mismatches between the fermionic populations before turning to the
normal unpaired state. For a wide range of values of such a mismatch
it turns out  that  the superfluid atoms tend to remain in the inner
core of the trap, whereas the normal atoms in excess  are expelled
 and form a shell surrounding the center of the trap.
These observations seem to support the  PS scenario   with phase
separation \cite{Bedaque:2003hi} between superfluid and normal
atoms. However,  a clear signature of the PS phase is still lacking
\cite{ketterle,ketterle2}.

The LOFF phase has not yet been  observed in experiments with
trapped cold atoms \cite{ketterle,Partridge}, although it was seen
in CeCoIn$_5$~\cite{radovan,NMR}. The portion of parameter space
where this phase is favored depends on the form of the condensate.
In the simplest case, in the weak coupling limit this region is
rather narrow and its detection in cold atom experiments might be
difficult \cite{ketterle}. However a definite conclusion can be
reached only after a detailed study of the space dependence of the
condensate and the knowledge of its dependence on the coupling
strength.

The present study represents a preliminary analysis of this problem.
We determine, in the mean field approximation, the phase diagram of
an interacting two species fermion system as a function of the
strength of the interaction and of their chemical potential
difference, paying special attention to the LOFF region. We choose a
four-fermion coupling to mimic the interaction, neglecting all the
effects of the trap in the fermionic Hamiltonian \footnote{For
preliminary studies of asymmetric spin populations that include the
effect of the trap see Ref.~\cite{trap}}. Due to these limitations
we do not attempt to make any comparison with experimental data,
leaving this task to future studies.

Other analyses have been presented already on the same subject
\cite{Pao,Son:2005qx,Sheehy}. In particular in Ref. \cite{Pao} a
similar study has been performed for the homogeneous case. We differ
from these authors for two reasons. First, we also include and
discuss the LOFF phase. Second, we work at fixed total density but
do not fix the difference in population between the two species.
Under this respect our study may be seen as complementary to Ref.
\cite{Pao}. Due to these assumptions, our phase diagram at $T=0$ is
given in terms of two parameters, $\delta\mu$ and the coupling
constant, similarly to the analysis presented in \cite{Son:2005qx}.

 One of our results is that the homogeneous phase
is energetically favored in a range of values of the chemical
potential difference that increases with increasing coupling
strength. Moreover it turns out that the phase transition between
the homogeneous and the LOFF phase is of the first order (this
result was already known in the weak coupling regime and we extend
it to the intermediate coupling regime) and there exists a critical
value of the coupling constant where the LOFF window shrinks to a
point. For larger values of the coupling  the LOFF phase cannot be
realized and the homogeneous phase has a first order phase
transition directly to the normal phase. Further increasing the
coupling strength and for large separations of the Fermi spheres the
excitation spectrum of the energetically stable homogeneous phase is
characterized by one gapless mode and the transition between the
superconductive phase and the normal phase becomes of the second
order.

As for the non-homogeneous phase we restrict our analysis to the
condensate characterized by a one-wave oscillation $\Delta({\bf x})
\sim \Delta~\exp{(i \bf{q \cdot x})}$. Non-homogeneous condensates
characterized by a more complicated space dependence, such as that
obtained by a superposition of several plane waves are expected to
produce  deeper minima of the free-energy $F$~\cite{Bowers:2002xr}.
Therefore our study only provides a lower bound for  the gap and the
window in $\delta\mu$ where the LOFF state is energetically favored.

The stability of the homogeneous phase with different Fermi surfaces
that we present in this article has been already discussed in
Ref.~\cite{Pao}. We will only consider the solutions of our system
of equations which are stable and that correspond to global minima
of the free-energy. We  do not examine the stability of the LOFF
phase which however is expected to be stable
\cite{Giannakis:2005sa}.

Our paper is organized as follows. In Section \ref{homogeneous} we
study the homogeneous superconductive phase constrained by the
conservation of the total number of fermions. In Section \ref{loff}
we discuss a LOFF phase characterized by a single plane-wave and we
determine the width of the LOFF window as a function of the
coupling. In Section \ref{conclusion} we summarize our results.

\section{Model and homogeneous phases \label{homogeneous}}
We shall analyze a system comprising fermions of two species, with
mass $m$, interacting via a short range (S-wave) four fermion
interaction. The Hamiltonian density is \be {\cal H} =
\sum_{\sigma=1,2}\bar\psi_\sigma(x) \left(-\frac{\nabla^2}{2 m } -
\mu_\sigma\right) \psi_\sigma(x)-G \bar\psi_1(x)
\bar\psi_2(x)\psi_2(x)\psi_1(x)\, ,\label{hamiltonian}\ee where
$G>0$ is the four fermion coupling constant. The chemical potentials
of the two species can be written as $\mu_1 = \mu + \delta\mu$ and
$\mu_2 = \mu - \delta\mu$, so that $\mu$ is the average of the two
chemical potentials and $2 \delta\mu$  their difference. We restrict
our analysis to zero temperature and work in the mean field
approximation.

 The effect of the attractive interaction between
fermions is to produce a gap $\Delta$ in the quasiparticle
dispersion laws which is related to the $\psi\psi$ condensate by\be
<\psi_{\alpha}(x) \psi_{\beta}(x)> = \frac{\Delta(x)}G\,
\epsilon_{\alpha\beta}\, . \ee The non-homogenous superconductive
phase, where the gap is not uniform in space, will be treated in the
next Section. Here we only discuss the homogeneous case,
$\Delta(x)=\Delta=$ const. In this case the excitation spectrum is
described by the quasiparticle dispersion laws \be \epsilon_1 =
+\delta\mu + E_p \,,~~~~~~~~ ~~~~\epsilon_2 = -\delta\mu    + E_p
\label{dispersionH} \, , \ee with \be E_p =
\sqrt{\xi^2+\Delta^2}\,,~~~~~~~ \xi= p^2/2m - \mu\,.\ee Using the
dispersion laws of the system one evaluates the grand-potential,
which is given, at $T=0$, by\be \Omega = \frac{\Delta^2}{G}
-\frac{1}{2}\int \frac{d^3 p}{(2 \pi)^3} \, \Big[|\epsilon_1| +
|\epsilon_2| - 2 \xi \Big] \label{omega}\, .\ee The integral in this
expression is ultraviolet divergent and can be regularized by the
usual procedure \cite{Melo}
 employing the S-wave
scattering length $a$:\be \frac{m}{4\pi a} = -\frac{1}G + m \int
\frac{d^3p}{(2\pi)^3} \frac{1}{p^2} \, . \label{slength} \ee We
introduce the dimensionless
 coupling constant \be g = \frac{1}{\pi k_F a}\, ,\ee
where $k_F$ is the Fermi momentum.  The weak coupling regime, where
the BCS approximation holds, corresponds to $g \to -\infty$. This
approximation is generally very good for superconductivity in
metals. On the other hand, in cold atoms the strength of the
interaction can be varied working in the vicinity of a Feshbach
resonance, where the scattering length strongly depends on the
applied magnetic field. Therefore both the weak and strong coupling
regimes can be reached in this case.

The gap parameter $\Delta$ and the mean chemical potential $\mu$ can
be determined self-consistently employing the equations \bea
\frac{\de \Omega}{\de \Delta} &=& 0 \label{gap} \, , \\
\frac{\de \Omega}{\de \mu} &=& -n  \label{number}\, ,\eea where
 $n = k_F^3/3 \pi^2$ is the fermionic number  density. Eqs. \eqref{gap} and \eqref{number}
 are the gap equation and the number equation respectively. In the weak coupling regime the chemical potential
$\mu$ differs from the Fermi energy $\epsilon_F$ by an amount of
order $\Delta^2/\mu^2$; therefore the approximation $\mu\simeq
\epsilon_F$ is usually adopted and the number equation is not used.

Let us note explicitly that we do not write equations for $\mu_1$
and $\mu_2$ separately. We work at fixed $n$, but do not impose
conditions on $\delta n=n_1-n_2$. As a consequence, we do not write
down a third equation: $\displaystyle\frac{\de \Omega}{\de\delta\mu}
= -\delta n$,  which would be needed in the analysis if $\delta n$
were held fixed \cite{Pao}. Since  $\delta n$ is not fixed,
conversions between particles of different species are allowed and
the thermodynamic potential whose minimum defines the vacuum state
is not $\Omega$, but $F=\Omega+\mu n$.

\subsection{Numerical solution of the self-consistent equations}
In order to determine the actual ground state of the system we solve
self-consistently Equations (\ref{gap}) and (\ref{number})  to
determine the values of $\Delta$ and $\mu$ for various values  of
the coupling $g$ and polarization $\delta\mu$.  From the computed
$\Delta$ and $\mu$ we evaluate  the free-energy $F = \Omega + \mu N$
in the superconductive phase which we compare  with the
corresponding result in the normal unpaired phase. For $\delta\mu=0$
this analysis was performed in Ref.~\cite{Engelbrecht} and, as a
test of our numerical code, we have reproduced all the results of
this paper. In particular we mention some results valid at
$\delta\mu=0$. The value of the gap is an increasing function of
$g$. The average chemical potential $\mu$ decreases and becomes
negative at $g \simeq 0.15$ signaling that the system has reached
the BEC phase. Finally, the superconductive phase is energetically
stable.

This qualitative behavior persists for moderate values of
$\delta\mu$, but increasing $\delta\mu$ eventually a transition to
the normal phase occurs. The nature of this transition we now
discuss.

For any fixed value of the coupling constant $g$ we denote by
$\delta\mu_c$ the largest chemical potential difference   that the
homogeneous superconductive phase can sustain. In other words for
$\delta\mu>\delta\mu_c$ the system enters the normal phase. Clearly
$\delta\mu_c$ depends on the coupling strength $g$. In the weak
coupling regime ($ g \to -\infty $) the value of the critical
polarization approaches the Clogston limit \cite{Clogston:1962}
$\delta\mu_c \simeq \Delta_0/\sqrt{2}$, where $\Delta_0$ is the
value of the gap at $\delta\mu=0$. The phase transition between the
superconductive and the normal phase is first order. With increasing
values of $g$ the value of $\delta\mu_c$ increases and the phase
transition remains first order for values of the coupling smaller of
$ \simeq 0.13$.

For values of the coupling in the range $0.13\lesssim g \lesssim
0.175$ the gap equation has three non trivial solutions in a range
of values of $\delta\mu$ close to the critical mismatch. One of
these solutions corresponds to a maximum of the free-energy, the
other two to local minima. The minima are favored for different
values of $\delta\mu$. At small values of $\delta\mu$ the favored
state is the one with $\Delta=\Delta_0$. For values of $\delta\mu$
larger than a critical value the favored state is the second one,
with $\Delta < \Delta_0$. The transition between these two states is
first order. We remark that such  behavior of the free-energy takes
place only in the range of the coupling $0.13 \lesssim g \lesssim
0.175$. For values  smaller than $\sim 0.13$ there is one phase
transition from the homogeneous to the normal phase. For values of
$g$ large than $\sim 0.175$ one of the minima of the free-energy
disappears and, increasing $\delta\mu$, one finds a second order
phase transition from the normal phase to the unpaired phase.

\begin{figure}[h!]
\vspace{0.5cm}
\includegraphics[width=3.in,angle=0]{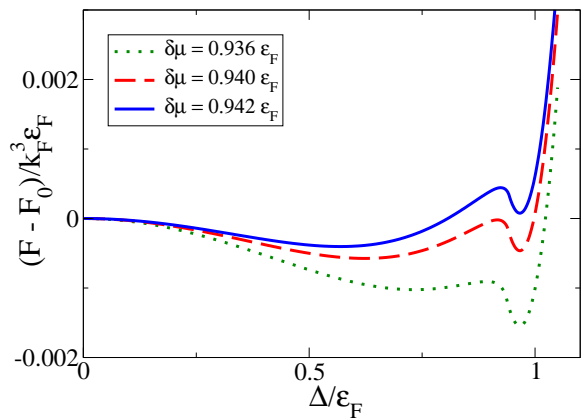}
\caption{Free energy difference $F-F_0$  as a function of $\Delta$
for various values of $\delta\mu$ at $g=0.135$.} \label{OmegaD}
\end{figure}

In order to clarify the  behavior in the above-mentioned range of
$g$, we plot in Fig.~\ref{OmegaD} the free-energy difference $F-F_0$
($F_0$ the value at $\Delta=0$) as a function of $\Delta$ for
various values of $\delta\mu$ at $g=0.135$, {\em i.e}. inside the
interval $[0.13,0.175]$. For each value of $\Delta$, the value of
$\mu$ is determined by the equation $\partial F/\partial\mu = 0$,
corresponding to Eq. \eqref{number}. We notice that, since the total
number density is fixed, the average chemical potentials of the
broken ($\Delta \neq 0$) and  normal ($\Delta=0$) phases are in
general different. For $\delta\mu = 0.936~\epsilon_F$ the
free-energy has a global minimum at $\Delta=\Delta_0\simeq
0.95~\epsilon_F$ and a local minimum at $\Delta \simeq
0.75~\epsilon_F$; at $\delta\mu= 0.940~\epsilon_F$ the two minima
are almost degenerate, and the values of the gap at the local minima
are $\Delta=\Delta_0$ and $\Delta\simeq 0.625~\epsilon_F$; finally
for $\delta\mu = 0.942~\epsilon_F$ the former local minimum becomes
the global one (and vice-versa), and the gap at the global minimum
is $\Delta\simeq 0.6~\epsilon_F$. For higher values of $\delta\mu$
the value of the gap decreases monotonically and for $\delta\mu =
\delta\mu_c \sim 0.955 ~\epsilon_F $ the system has a second order
phase transition to the normal phase.

\begin{figure}[!t]
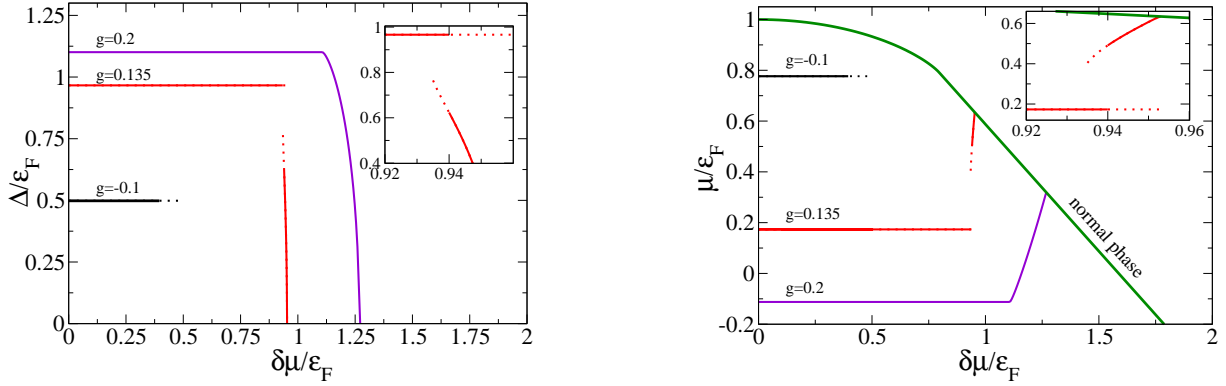
\vspace{0.5cm}
\includegraphics[width=7cm]{gap.eps}\hskip2cm~\includegraphics[width=7cm]{mug.eps}
\caption{On the left: The gap $\Delta/\epsilon_F$ vs.
$\delta\mu/\epsilon_F$ for three values of the dimensionless
coupling $g$. From top to bottom the lines refer to $g=0.2$ (purple
online), $g=0.135$ (red online) and $g=-0.1$ (black online). The $
g=0.2$
 curve shows a second order phase transition to the normal phase at
 $\delta\mu/\epsilon_F\simeq 1.27$. The $g=-0.1$ curve shows a
 first-order transition to the normal phase at $\delta\mu\simeq0.39 \epsilon_F$. The
intermediate curve ($g=0.135$), shown in more detail in the inset,
shows the existence of two phase transitions. One phase transition
is first-order. It leads to a superconductive phase with a
different, smaller, value of the gap. The second transition leads
smoothly to the normal phase.  On the right: $\mu/\epsilon_F$ vs.
$\delta\mu/\epsilon_F$ for the same three values of the
dimensionless coupling $g$.  The continuous upper curve (green
online) refers to the normal phase ($g\to-\infty$). The other three
curves from bottom to top refer to $g=0.2$ (purple online),
$g=0.135$ (red online) and $g=-0.1$ (black online). The inset
represents an enlargement of the curve at $g=0.135$. In both panels
the dotted parts of the $g=0.135$ and of the $g=-0.1$ lines
correspond to metastable states.} \label{Deltafig}
\end{figure}

The dependence on $g$ of the order of the phase transitions is shown
on the left panel in Fig.~\ref{Deltafig} by three representative
values of the dimensionless coupling constant, one inside the
interval $[0.13,0.175]$, another one on the left, and a third one on
the right of the interval. The lowest curve refers to $g=-0.1$. We
have not considered here the possibility of inhomogeneous
superconductivity and therefore we have a first order phase
transition from the superconductive to the normal state. It occurs
at $\delta\mu \simeq 0.79 \Delta_0$. For   $0.79 \lesssim
\delta\mu/\Delta_0 \leq 1$ the superconductive phase becomes
metastable and is shown  as a dotted  line.  The highest curve is
computed at $g=+0.2$: for this value the transition from the
superconductive to the normal phase is second order. The
intermediate curve is obtained at $g=+0.135$ and shows, in agreement
with the results of Fig. \ref{OmegaD}, two phase transitions: a
first order phase transition from the value 0.95 to the value 0.65
of the gap parameter, and a second order phase transition to the
normal phase. The values corresponding to the metastable phases are
depicted as dotted curves. An enlarged picture of this case is in
the inset. On the right panel in Fig.~\ref{Deltafig} we show the
behavior of the average chemical potential $\mu$ as a function of
$\delta\mu$, for the same values of the dimensionless coupling
constant $g$. In the figure, the  upper curve (green online)
represents the average chemical potential in the normal phase The
inset refers again to $g=0.135$.

\begin{figure}[h!]\vspace{0.5cm}
\includegraphics[width=3.in,angle=0]{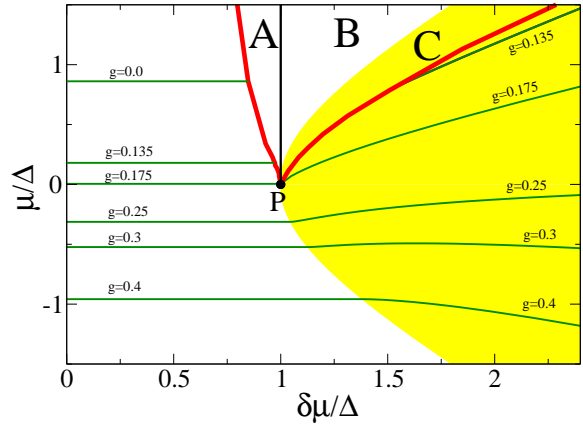}
\caption{Full (green online) lines are simultaneous solutions of the
gap and number equations  for different values of the coupling
constant and of the mismatch between the Fermi spheres. The regions
above  the full (red online) line labeled with {\bf A}, {\bf B} and
{\bf C} correspond to phases where no physical solutions of the gap
and number equations have been found. The  phase in region {\bf B},
the Breached Pairing phase, corresponds to  unstable solutions.
Regions {\bf A} and {\bf C} correspond to metastable phases. The
shadow (yellow online) region, with the exclusion of the region {\bf
C}, corresponds to the region with stable gapless solutions at one
sphere in momentum space. The remaining part of the diagram
corresponds to stable gapped solutions. } \label{sdreus}
\end{figure}

It is also worth mentioning that the first order phase transition
between the two minima of the free-energy corresponds to a phase
transition between a gapped and gapless phase. The gapless phase is
characterized by having one zero in the  quasiparticle dispersion
law at one sphere in momentum  space. Had the  dispersion laws two
zeros then the system could live in the Breached Pairing phase
\cite{Liu:2002gi,Gubankova:2003uj}, but this possibility is not
realized in this model at least within the present approximations.
To illustrate this point we have reported in Figure~\ref{sdreus} the
results for $\mu/\Delta$ vs. $\delta\mu/\Delta$ as lines (green
online) labeled with various values of $g$.  Since for some values
of $g$ there are first order phase transitions, some regions of this
diagram are never reached by stable physical states, which is why in
some cases the lines are interrupted. Such regions are above the
thick full (red online) line, which has been determined comparing
the energies of the various phases, and have been labeled with the
letters {\bf A}, {\bf B} and {\bf C}. The regions labeled as {\bf A}
and {\bf C} correspond to metastable points that are local minima of
the free-energy. For $g=0.135$ they were reported in the insets of
Fig. \ref{Deltafig} as dotted points in the upper curve and lower
curve respectively. The points in region labeled as {\bf B}
correspond to unstable BP points  that are maxima of the
free-energy. The remaining parts of the diagram correspond to
allowed regions. The white area corresponds to the stable gapped
phase and the shadow area (yellow online), with the exclusion of the
region {\bf C}, to the stable gapless superconductive phase. In the
shadow  region $\delta\mu > \sqrt{\mu^2 + \Delta^2}$   there are
gapless excitations at one sphere in momentum space. All the regions
meet at the point {\bf P}, on the line corresponding to $g=0.175$.
The meaning of this point will be clarified below.

\section{LOFF Phase \label{loff}}
For values of $\delta\mu$ larger than $\delta\mu_c$ it can be
energetically convenient for the fermionic  system to form  Cooper
pairs with non-zero total momentum \cite{LOFF}.  In the following we
will consider a simple non homogeneous LOFF condensate characterized
by a single plane wave\be <\psi_{\alpha}(x) \psi_{\beta}(x)> =
\Delta \epsilon_{\alpha\beta} e^{i 2 \bf q \cdot x}
\,,\label{ansatz}\ee where ${\bf 2q}$ is the total momentum of the
pair. More complicated patterns, such as those arising  by more
plane waves, may lead to states with a lower free-energy, but for
our illustrative purposes the ansatz \eqref{ansatz} is sufficient.

In the LOFF phase the quasiparticle dispersion laws are given by
\be\epsilon_1({\bf q}) = +\delta\mu - \frac{{\bf q \cdot p}}m +
\sqrt{\xi(q)^2 + \Delta^2} \, ,~~~~~~~~ \epsilon_2({\bf q}) =
-\delta\mu + \frac{{\bf q \cdot p}}m  + \sqrt{\xi(q)^2 + \Delta^2}
\, ,\label{dispersionL}\ee where we have defined
$\displaystyle\xi(q) = \frac{p^2+q^2}{2m} - \mu$. The free-energy
for the LOFF phase can be written as \be\Omega(q) =
\frac{\Delta^2}{G} -\frac{1}{2}\int \frac{d^3 p}{(2 \pi)^3} \,
\Big[|\epsilon_1({\bf q})| + |\epsilon_2({\bf q})| - 2 \xi( q) \Big]
\label{omegaLOFF}\, ,\ee and the integral is regulated as in the
homogeneous case, i.e. by employing the S-wave scattering length
defined in Eq.(\ref{slength}). In addition to the gap and number
Equations (\ref{gap}),(\ref{number}) one has to consider the
equation \be\frac{\de \Omega}{\de q} = 0 \label{qequation}\, ,\ee
which determines  the modulus of $\bf q$; the direction of  $\bf q$
is spontaneously determined by the system. Therefore in the
non-homogeneous phase one has to solve a system of three coupled
equations: the number equation, the gap equation   and
Eq.(\ref{qequation}) for $\mu$, $\Delta$ and $q$ as a function of
$g$ and $\delta\mu$ and to look for minima of the free-energy $F =
\Omega + \mu n $.

\begin{figure}[!th]\vspace{0.5cm}
\includegraphics[width=2.in,angle=-90]{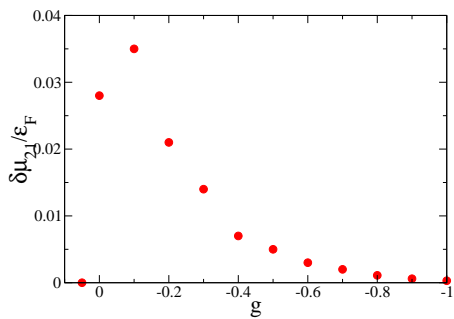}
\caption{Amplitude of the LOFF window $\delta\mu_{21} =
\delta\mu_2-\delta\mu_1$ as a function of the coupling $g$. The LOFF
window has its maximum amplitude for $g\simeq -0.1$ and shrinks to
zero for $g > 0.05$. } \label{window}
\end{figure}

From  general arguments \cite{Casalbuoni:2003wh} one knows that the
one-wave LOFF phase is energetically favored for values of the
mismatch $\delta\mu$ in some interval:
$\delta\mu_1<\delta\mu<\delta\mu_2$. Let us  define the amplitude of
the LOFF window as $\delta\mu_{21} = \delta\mu_2 - \delta\mu_1$.
Since the free-energy in the one-wave LOFF phase differs only
slightly from the value in the normal phase, it turns out that
$\delta\mu_1 \simeq \delta\mu_c$. In the previous Section we have
found that $\delta\mu_c$ is an increasing function of $g$. Comparing
 the free-energy of the LOFF phase with the free-energy of the
normal phase we obtain that  also $\delta\mu_2$ is an increasing
function of $g$ but $\delta\mu_{21}$ turns out to be  a non
monotonic function of the coupling. In the weak coupling regime we
recover the well known result $\delta\mu_2 \simeq 0.754 \Delta_0 $
and therefore $\delta\mu_{21}=(0.754-0.707)\Delta_0=0.047 \Delta_0 $
is an increasing function of $g$. However in the intermediate
coupling regime $\delta\mu_1$ grows faster than $\delta\mu_2$ and
the amplitude of the LOFF window, depicted in Fig.\ref{window},
reaches its maximum amplitude at $g \simeq -0.1$ and then begins to
shrink; for $g \simeq 0.05$ the LOFF phase disappears.

The largest value of the LOFF window  is about $0.035 \,
\epsilon_F$. However, as we have already mentioned, this value is a
lower bound for $\delta\mu_{21}$ because more complicated
crystalline structures might be able to sustain larger values of the
mismatch between the Fermi surfaces.

As a check of our results, we have determined the second order phase
transition lines of the phase diagram by a Ginzburg-Landau (GL)
expansion of the grand potential $\Omega$, both in the homogeneous
and in the LOFF phase. The use of this approximation is justified
because also in the strong coupling regime one has $\Delta/\delta\mu
\to 0$ near the second order lines. Since we are interested to the
second order transitions, it is enough to expand $\Omega$ up to the
fourth order in $\Delta$, so the grand potential can be written as
\begin{equation}
\Omega = \Omega_0 + \frac{\alpha}{2}\Delta^2 +
\frac{\beta}{4}\Delta^4~, \label{eq:GLomega}
\end{equation}
where $\Omega_0$ is the free gas contribution and the coefficients
are given by
\begin{subequations}
\begin{equation}
\alpha = \frac{2}{G} +
T\!\sum_{n=-\infty}^{\infty}\int\frac{d^3p}{(2\pi)^3}\frac{2}{(i\omega_n-\epsilon_1)(i\omega_n+\epsilon_2)}~,
\end{equation}
\begin{equation}
\beta =
T\!\sum_{n=-\infty}^{\infty}\int\frac{d^3p}{(2\pi)^3}\frac{2}{(i\omega_n-\epsilon_1)^2(i\omega_n+\epsilon_2)^2}~.
\end{equation}\label{eqs:coeffGL}
\end{subequations}
In Eqs.~\eqref{eqs:coeffGL} the $\epsilon_\sigma$ are the dispersion
laws of the quasi-particles,
\begin{equation}
\epsilon_1 = \frac{({\bf p}+{\bf q})^2}{2m} - \mu_1~,~~~~~
\epsilon_2 = \frac{({\bf p}-{\bf q})^2}{2m} - \mu_2
\end{equation}
(the homogeneous case is studied by putting ${\bf q}=0$ in the above
expressions). The divergence in the integral defining  the
coefficient $\alpha$ is  cured, by the introduction of the S-wave
scattering length, as discussed in Section \ref{homogeneous}. Using
the GL expansion we reproduce within a few percent the second order
transition lines obtained by the numerical evaluation of the
free-energy minima in the full theory .

\section{Phase Diagram \label{conclusion}}
\begin{figure}[!th]
\includegraphics[width=3.in,angle=0]{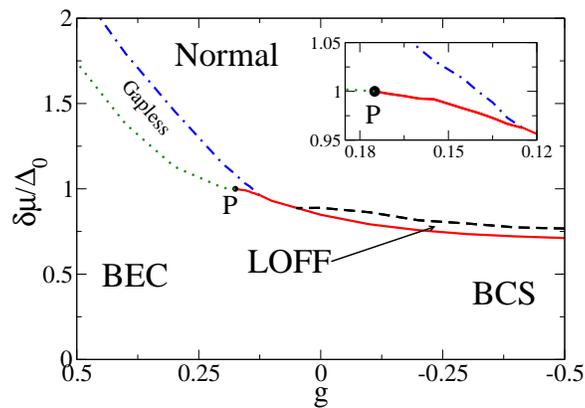}
\caption{Phase-diagram at $T=0$. The full line (red online)
indicates the first order phase transition between the homogeneous
gapped superfluid phase and the LOFF phase for $g\lesssim 0.05$ or
the normal phase for $0.05 \lesssim g\lesssim 0.13$ or the gapless
homogeneous superfluid phase  for $0.13 \lesssim g\lesssim 0.175$.
The dashed line (black online) indicates the second order phase
transition between the LOFF phase and the normal phase. The
dot-dashed line (blue online) indicates  the second order phase
transition between the homogeneous superconductive phase and the
normal phase. The dotted (green online) line, which does not
correspond to a phase transition, separates the homogeneous  gapped
phase from the homogeneous gapless phase.  In the inset it is shown
that the full line continues beyond the point where the
 dot-dashed  line and the full  line meet.} \label{phasediagram}
\end{figure}

 We summarize our results in the phase diagram depicted  in Fig.
 \ref{phasediagram}. In the following discussion of the phase
 diagram we will show that there is a correspondence between some
 regions and lines of the phase diagram and of the diagram depicted in
 Fig.~\ref{sdreus}.

To begin with we describe the   full line (red online).  At  values
of $g\lesssim0.05$ it  indicates the first order phase transition
between the homogeneous superconductive phase and the LOFF phase.
For $0.05\lesssim g\lesssim 0.13$ the full line indicates a first
order phase transition from the superconductive phase to the normal
phase. The regions corresponding to the LOFF phase and to the normal
phase are not depicted in Fig.~\ref{sdreus}.

For $0.13 \lesssim g \lesssim 0.175$, the full line represents the
first order phase transition between the superfluid gapped phase and
the superfluid gapless phase (see  the discussion in Section
\ref{homogeneous} in connection with Figs. \ref{OmegaD} and
\ref{Deltafig}).  From the inset of Fig.~\ref{phasediagram} one can
see that such a first order phase  transition line terminates at the
point P, at $g\sim 0.175$. This
 point is  depicted in Fig.~\ref{sdreus}, by the same letter.
The full (red) line shown Fig.~\ref{phasediagram}  corresponds to
the full (red) line shown in Fig.~\ref{sdreus}. One can understand
the different topologies of Figs.~\ref{phasediagram} and
\ref{sdreus} in the following way. Since the regions {\bf A}, {\bf
B} and {\bf C} in Fig.~\ref{sdreus} are not physical they cannot be
reported in the phase diagram of Fig.~\ref{phasediagram}. One can
naively say that in the phase diagram   the areas of regions {\bf
A}, {\bf B} and {\bf C}   have been shrunk to zero. Therefore once
the two branches of the full (red) line  of Fig.~\ref{sdreus} are
reported in Fig.~\ref{phasediagram} they overlap as one line.

For larger values of $g \gtrsim 0.13$ we do not find a phase
transition between the gapped and the gapless phase which are
separated in Fig.~\ref{phasediagram} by the dotted (green online)
line. Such a line corresponds  to the onset of the shadow (yellow
online) region of Fig.~\ref{sdreus} for negative values of $\mu$.
The shadow region (with the exception of the region {\bf C}) shown
in Fig.~\ref{sdreus} corresponds to  the region between the dotted
green line and the dot-dashed line ( blue online) of
Fig.~\ref{phasediagram}. The dot-dashed line represents the second
order phase transition between the superfluid phase and the normal
phase.

The phase transition between the LOFF phase and the normal phase
(dashed red line) is  of the second order. This result, already
known in the weak coupling approximation~\cite{LOFF}, persists also
 for  intermediate  values of the dimensionless coupling constant,
{\em i.e.} up to $g\sim+0.05$. For this value of $g$ and for
$\delta\mu \sim 0.8 \epsilon_F$,  the transition lines LOFF-BCS and
LOFF-Normal phase cross. For values of the coupling constant larger
than  $g \sim 0.05$ the LOFF phase cannot be realized, see the
discussion related to Fig. \ref{window}.

In conclusion we have analyzed in the mean field approximation the
phase diagram of a fermion superfluid system comprising two
unbalanced populations. We have worked  at fixed total number
density but arbitrary density difference, conversions between
particles of different species are allowed thereof. As a
consequence, at $T=0$, the phase diagram depends on two parameters,
the dimensionless coupling constant and the mismatch $\delta\mu$
between the two Fermi spheres.

In view of possible applications to the study of the phase diagrams
of systems composed of cold fermion atoms the present study should
be considered as preliminary. As a matter of fact we have not
included two effects. First, we have included only fermion
self-interactions, without taking into account a confining
potential. This development is postponed to future investigations.
Second, we have not included fluctuations around the mean field
solution. In the literature this is considered in general as a good
approximation, at $T=0$ \cite{Engelbrecht}. However, as discussed in
\cite{Carlson:2005}, for values of the coupling smaller than some
critical value (with our definitions, this corresponds to the region
$|g|<0.1$) one enters a region where fluctuations may play a role.
In general including fluctuations has the effect to enlarge the
region of the ordered phase. This is revealed in the Quantum Monte
Carlo computation of \cite{Carlson:2005} by a gap in the dispersion
law of the quasiparticle in a region where the normal phase should
already be present. In other words the effect of the fluctuations
would be to increase the transition line around $g\sim 0$ in Fig.
\ref{phasediagram}. Therefore it is plausible that a more advanced
study, would reveal in the region $|g|<0.1$ a more complex
structure, more akin to the one proposed in \cite{Son:2005qx}. We
plan to come to this problem in the future.

\section{Acknowledgement}
We would like to thank M.~Alford, R.~Casalbuoni, M.~Ciminale,
E.~Gubankova, V.~Laporta, K.~Rajagopal, A.~Schmitt and R.~Sharma,
 for useful discussions. One of us (M.R.)
would like to thank the Center for Theoretical Physics of MIT for
kind hospitality. The work of MM has been supported by the ``Bruno
Rossi" fellowship program and by the U.S. Department of Energy
(D.O.E.) under cooperative research agreement \#DE-FC02-94ER40818.

\end{document}